\newif\ifdraft
\draftfalse

\newif\ifcameraready
\camerareadytrue

\documentclass[camera,letterpaper,nomarginnotes,nonarrowgutter]{jpaper}
\newcounter{version}
\usepackage{mathptmx} %
\usepackage[sort,compress]{cite}
\usepackage{amsmath,amssymb,amsfonts}
\usepackage{algorithmic}
\usepackage{graphicx}
\usepackage{textcomp}
\usepackage{xcolor}
\usepackage{balance}
\usepackage{amsmath,amssymb,amsfonts}
\usepackage{graphicx}
\usepackage{textcomp}
\usepackage{pifont}
\usepackage{booktabs}
\usepackage{tabularx}
\usepackage{glossaries} %
\usepackage{setspace}
\usepackage{listings}
\usepackage[linesnumbered]{algorithm2e}
\usepackage{siunitx} %
\usepackage{subcaption}
\usepackage{tikz}
\usepackage{xcolor}
\usepackage{multirow}
\usepackage{enumitem}

\usepackage{titlesec}

\usepackage{interval} %
\usepackage{duckuments} %
\usepackage{algorithmic} %

\usepackage[us,12hr]{datetime}
\usepackage[en-GB, useregional=numeric]{datetime2}
\usepackage{fourier-orns}
\usepackage{fancyhdr}
\usepackage{pgfornament}
\usetikzlibrary{calc}
\usepackage{url}

\newcommand{\tras}[0]{t_{RAS}}
\newcommand{\trp}[0]{t_{RP}}
\newcommand{\trc}[0]{t_{RC}}
\newcommand{\trefi}[0]{t_{REFI}}
\newcommand{\trefw}[0]{t_{REFW}}
\newcommand{\trfc}[0]{t_{RFC}}
\newcommand{\trrd}[0]{t_{RRD}}

\newcommand{\nrg}[0]{$N_{RG}$}

\newcommand{\act}[0]{\texttt{ACT}}
\newcommand{\pre}[0]{\texttt{PRE}}
\newcommand{\refresh}[0]{REF}
\newcommand{\wri}[0]{\texttt{WR}}
\newcommand{\rd}[0]{\texttt{RD}}

\newacronym{acmin}{$AC_{min}$}{the minimum number of total aggressor row activations to cause at least one bitflip}

\newcommand{\apa}[0]{\texttt{APA}}

\newcommand{\vddh}[0]{\texttt{$V_{DD}/2$}}

\newcommand{\pum}[0]{PUM}
\newcommand{\pim}[0]{PIM}
\newcommand{\pnm}[0]{PNM}

\newcommand{\pud}[0]{{PUD}}

\usepackage[shortcuts]{extdash}

\newacronym{iqr}{$IQR$}{inter-quartile range}
\newacronym{act}{\act{}}{activate}
\newacronym{pre}{\pre{}}{precharge}
\newacronym{ref}{\refresh{}}{refresh}
\newacronym{wr}{\wri{}}{write}
\newacronym{rd}{\rd{}}{read}
\newacronym{pim}{\pim{}}{Processing-In-Memory}
\newacronym{pnm}{\pnm{}}{Processing-Near-Memory}

\newacronym{pum}{\pum{}}{Processing-Using-Memory}
\newacronym{pud}{\pud{}}{Processing-Using-DRAM}
\newacronym{apa}{\apa{}}{\act{} $\rightarrow$ \pre{} $\rightarrow$ \act{}}
\newacronym{jedec}{JEDEC}{Joint Electron Device Engineering Council}
\newacronym{nrg}{\nrg{}}{a group of N rows activated simultaneously with \apa{} sequence as a row group}

\newacronym{trefw}{$\trefw$}{refresh window}
\newacronym{tras}{$\tras$}{the latency of sensing the row's data and fully restoring a DRAM cell's charge}
\newacronym{trp}{$\trp$}{the latency of de-asserting a wordline and precharging the bitlines to \vddh{}}
\newacronym{trc}{$\trc$}{the minimum time needed between two consecutive row activations targeting the same bank}
\newacronym{trefi}{$\trefi$}{refresh interval}
\newacronym{trfc}{$\trfc$}{refresh latency}
\newacronym{trrd}{$\trrd$}{the minimum time needed between two consecutive row activations targeting the same rank}
\newacronym{puf}{PUF}{physical unclonable function}
\newacronym{trn}{TRN}{true random number}

\newcommand{\revdel}[1]{}

\definecolor{gfored}{rgb}{0.580, 0.050, 0.211}
\definecolor{ao}{rgb}{0.007, 0.520, 0.867}
\definecolor{moegi}{rgb}{0.357, 0.537, 0.188}
\definecolor{jl}{rgb}{1.0, 0.2, 0.8}
\definecolor{brown(web)}{rgb}{0.65, 0.16, 0.16}
\definecolor{bisque}{rgb}{1.0, 0.89, 0.77}
\definecolor{nbs}{rgb}{0.88, 0.07, 0.37}
\definecolor{yt}{rgb}{0.58, 0.44, 0.86}
\definecolor{iy}{rgb}{0.0, 0.36, 0.05}
\definecolor{mel}{rgb}{0.9, 0.55, 0.31}

\newcommand{\dingOne}{\circledtest{1}}
\newcommand{\dingTwo}{\circledtest{2}}
\newcommand{\dingThree}{\circledtest{3}}

\newcommand{\xxx}[1]{\param{XXX}} %

\newcommand{\ignore}[1]{}

\newcommand{\param}[1]{\textcolor{red}{#1}}

\usepackage[colorinlistoftodos,prependcaption,textsize=scriptsize]{todonotes} %
\usepackage{marginnote} 
\let\marginpar\marginnote
\newcommand{\revAp}[1]{}

\newcommand{\revBp}[1]{}

\newcommand{\revDp}[1]{}

\newcommand{\revEp}[1]{}

\newcommand{\atb}[1]{{#1}}
\newcommand{\atbcomment}[1]{}
\newcommand{\om}[1]{{#1}}

\ifdraft
    \usepackage[colorinlistoftodos,prependcaption,textsize=tiny]{todonotes}
    \usepackage{color,soul}

    \newcommand{\agycomment}[1]{\todo[size=\scriptsize, linecolor=orange, bordercolor=orange, backgroundcolor=white]{\textcolor{gfored}{\textbf{@gy:} #1}}}
    \renewcommand{\agycomment}[1]{{\textcolor{gfored}{\textbf{\hl{[@gy:}} \hl{#1}\textbf{]}}}}

    \newcommand{\mscomment}[1]{\todo[size=\scriptsize, linecolor=orange, bordercolor=orange, backgroundcolor=white]{\textcolor{red}{\textbf{@MS:} #1}}}
    \renewcommand{\mscomment}[1]{{\textcolor{red}{\textbf{\hl{[@MS:}} \hl{#1}\textbf{]}}}}

    \newcommand{\atb}[1]{\textcolor{ao}{#1}}

    \newcommand{\atbcomment}[1]{\todo[size=\scriptsize, linecolor=orange, bordercolor=orange, backgroundcolor=white]{\textcolor{ao}{\textbf{@atb:} #1}}}

    \newcommand{\yctcomment}[1]{\todo[size=\scriptsize, linecolor=orange, bordercolor=orange, backgroundcolor=white]{\textcolor{yt}{\textbf{@yct:} #1}}}

    \newcommand{\gfcomment}[1]{\todo[size=\scriptsize, linecolor=orange, bordercolor=orange, backgroundcolor=white]{\textcolor{blue}{\textbf{@gf:} #1}}}

    \newcommand{\nbcomment}[1]{\todo[size=\scriptsize, linecolor=orange, bordercolor=orange, backgroundcolor=white]{\textcolor{nbs}{\textbf{@nb:} #1}}}

    \newcommand{\hluocomment}[1]{\todo[size=\scriptsize, linecolor=orange, bordercolor=orange, backgroundcolor=white]{\textcolor{moegi}{\textbf{@hluo:} #1}}}

    \newcommand{\melcomment}[1]{\todo[size=\scriptsize, linecolor=red, bordercolor=red, backgroundcolor=white]{\textcolor{mel}{\textbf{@mel:} #1}}}

    \newcommand{\ieycomment}[1]{\todo[size=\scriptsize, linecolor=orange, bordercolor=orange, backgroundcolor=white]{\textcolor{iy}{\textbf{@iey:} #1}}}

    \newcommand{\omcomment}[1]{\todo[size=\scriptsize, linecolor=orange, bordercolor=orange, backgroundcolor=white]{\textcolor{teal}{\textbf{@ste:} #1}}}

\else
    \renewcommand{\param}[1]{\textcolor{black}{#1}}

    \newcommand{\agycomment}[1]{}
    \newcommand{\agyinline}[1]{}

    \newcommand{\mscomment}[1]{}

    \newcommand{\yctcomment}[1]{}

    \newcommand{\gfcomment}[1]{}

    \newcommand{\nbcomment}[1]{}

    \newcommand{\hluocomment}[1]{}

    \newcommand{\melcomment}[1]{}

    \newcommand{\ieycomment}[1]{}
    
    \newcommand{\omcomment}[1]{}

\fi

\newcommand*\DRAMCMD[1]{\texttt{#1}}
\newcommand*\DRAMTIMING[1]{t\textsubscript{#1}}

\definecolor{frenchblue}{rgb}{0.19, 0.55, 0.91}

\usepackage[most]{tcolorbox} 
\tcbset{before skip=4.1pt, after skip=4.1pt}

\newtcolorbox[auto counter]{obsx}[3][]{%
    colframe = #2!45,
    colback  = #2!10,
    coltitle = #2!20!black, 
    colbacktitle=#2!20,
    coltitle=black,
    fonttitle=\bfseries, 
    title=#3~\thetcbcounter.\ ,
    enhanced,
    attach boxed title to top left={yshift=-2.8mm, xshift=0.15cm},
    bottom=-2.2pt,
    #1%
}

\usepackage[most]{tcolorbox} 
\newtcolorbox[auto counter]{tkx}[2][]{%
    enhanced, breakable, center title,
    colframe = #2!45,
    colback  = #2!10,
    colbacktitle=#2!20,
    left=-0.5pt,
    right=-0.5pt,
    bottom=-2pt,
    top=-0.25pt,
    #1%
}

\newcounter{obs}
\newcommand\observation[1]{%
   \refstepcounter{obs}
  \vspace{0.3em}
  \noindent
  \begin{tabular}{|p{0.95\linewidth}|}
       \hline
       \textbf{{Observation \theobs}.} {{#1}}\\
       \hline 
  \end{tabular}
    \vspace{0.05em}
}
\definecolor{whitesmoke}{rgb}{0.96, 0.96, 0.96}

\newcounter{tkw}
\newcommand\takeawaybox[1]{%
   \refstepcounter{tkw}
   \vspace{0.3em}

  \noindent
  \begin{tabular}{|p{0.95\linewidth}|}
       \hline
       \rowcolor{whitesmoke}
       \textbf{{Takeaway \thetkw}.} {{#1}}\\
       \hline
  \end{tabular}
    \vspace{0.05em}

}

\newcounter{hypo}
\newcommand\hypobox[1]{%
   \refstepcounter{hypo}
   \vspace{0.3em}

  \noindent
  \begin{tabular}{|p{0.95\linewidth}|}
       \hline
       \rowcolor{whitesmoke}
       \textbf{{Hypothesis \thehypo}.} {{#1}}\\
       \hline
  \end{tabular}
    \vspace{0.05em}

}

\newcommand{\exploitingRowHammerAllCitationsExceptFlipping}[0]{\cite{rowhammer-js,  fournaris2017exploiting, poddebniak2018attacking, tatar2018throwhammer, carre2018openssl, barenghi2018software, zhang2018triggering, bhattacharya2018advanced, google-project-zero, rowhammergithub, seaborn2015exploiting, van2016drammer, gruss2016rowhammer, razavi2016flip, pessl2016drama, xiao2016one, bosman2016dedup, bhattacharya2016curious, burleson2016invited, qiao2016new, brasser2017can, jang2017sgx, aga2017good, mutlu2017rowhammer, tatar2018defeating, gruss2018another, lipp2018nethammer, van2018guardion, frigo2018grand, cojocar2019eccploit,  ji2019pinpoint, mutlu2019rowhammer, hong2019terminal, kwong2020rambleed, frigo2020trrespass, cojocar2020rowhammer, weissman2020jackhammer, zhang2020pthammer, yao2020deephammer, deridder2021smash, hassan2021utrr, jattke2022blacksmith, tol2022toward, kogler2022half, orosa2022spyhammer, zhang2022implicit, liu2022generating, cohen2022hammerscope, zheng2022trojvit, fahr2022frodo, tobah2022spechammer, rakin2022deepsteal, mus2023jolt, tol2023dontknock, mutlu2023fundamentally, jattke2024zenhammer}}

\newcommand{\figref}[1]{Fig.~\ref{#1}}

\newcommand*\circledtest[1]{\tikz[baseline=(char.base)]{
            \node[shape=circle,fill,inner sep=0.3pt] (char) {\textcolor{white}{#1}};}}

\ifcameraready

    \newcommand{\atbcr}[2]{\ifnum#1>-1\textcolor{black}{#2}\else{#2}\fi}
    \newcommand{\ieycr}[2]{\ifnum#1>-1\textcolor{black}{#2}\else{#2}\fi}
    \newcommand{\omcr}[2]{\ifnum#1>-1\textcolor{black}{#2}\else{#2}\fi}
    \newcommand{\omcrcomment}[1]{}
    \newcommand{\ominline}[1]{}
    \newcommand{\ieycrcomment}[1]{}
    \newcommand{\ieyinline}[1]{}

\else
    
    \newcommand{\atbcr}[2]{\ifnum#1=\value{version}\textcolor{ao}{#2}\else{#2}\fi}

    \newcommand{\ieycr}[2]{\ifnum#1=\value{version}\textcolor{blue}{#2}\else{#2}\fi}

    \newcommand{\ieycrcomment}[1]{\todo[linecolor=orange, bordercolor=orange, backgroundcolor=white]{\textcolor{iy}{\textbf{@Ismail:} #1}}}
    \newcommand{\ieyinline}[1]{\\\textcolor{iy}{\textbf{@Ismail:} #1}}

    \newcommand{\omcr}[2]{\ifnum#1=\value{version}\textcolor{blue}{#2}\else{#2}\fi}
    \newcommand{\omcrcomment}[1]{\todo[linecolor=orange, bordercolor=orange, backgroundcolor=white]{\textcolor{red}{\textbf{@Onur:} #1}}}
    \newcommand{\ominline}[1]{\\\textcolor{red}{\textbf{@Onur:} #1}}

\fi

\makeatletter
\g@addto@macro{\normalsize}{%
 \setlength{\abovedisplayskip}{2pt plus 1pt minus 1pt}
 \setlength{\belowdisplayskip}{2pt plus 1pt minus 1pt}
  \setlength{\abovedisplayshortskip}{0pt}
  \setlength{\belowdisplayshortskip}{0pt}
  \setlength{\intextsep}{2pt plus 1pt minus 1pt}
  \setlength{\textfloatsep}{2pt plus 1pt minus 1pt}
  \setlength{\skip\footins}{5pt plus 1pt minus 1pt}}
\makeatother

\usepackage{tikz}
\usetikzlibrary{arrows}

\usepackage{lscape}
\usepackage[bookmarks=true,breaklinks=true,colorlinks,linkcolor=black,citecolor=blue,urlcolor=black]{hyperref}

\def\BibTeX{{\rm B\kern-.05em{\sc i\kern-.025em b}\kern-.08em
    T\kern-.1667em\lower.7ex\hbox{E}\kern-.125emX}}
    
\newcommand{\versionnum}[0]{3.0}

\ifcameraready
    \fancypagestyle{plain}{
    \fancyhead{}
    }
\else
    \fancypagestyle{firstpage}
    {
    }
\fi

\makeatletter
\def\bstctlcite{\@ifnextchar[{\@bstctlcite}{\@bstctlcite[@auxout]}}
\def\@bstctlcite[#1]#2{\@bsphack
  \@for\@citeb:=#2\do{%
    \edef\@citeb{\expandafter\@firstofone\@citeb}%
    \if@filesw\immediate\write\csname #1\endcsname{\string\citation{\@citeb}}\fi}%
  \@esphack}
\makeatother

\begin{document}
\bstctlcite{IEEEexample:BSTcontrol}

\title{An Experimental Characterization of\\Combined RowHammer and RowPress Read Disturbance\\in Modern DRAM Chips}
\vspace{-.5em}
\author{
Haocong~Luo\quad
{\.I}smail~Emir~Y{\"u}ksel\quad
Ataberk~Olgun\quad
A.~Giray~Ya\u{g}l{\i}k\c{c}{\i}\quad\\
Mohammad~Sadrosadati~\quad
Onur~Mutlu\vspace{1.5pt}\\
\emph{ETH~Z{\"u}rich}~\qquad
\vspace{-.5em}
}

\maketitle
\ifcameraready
    \thispagestyle{plain}
\else
    \thispagestyle{firstpage}
\fi

\pagestyle{plain}

\setcounter{version}{3}
\begin{abstract}
DRAM read disturbance can break memory isolation, a fundamental property to ensure system {robustness (i.e., reliability, security, safety)}. RowHammer and RowPress are two different DRAM read disturbance phenomena. RowHammer induces bitflips in physically adjacent {victim} DRAM rows by repeatedly opening and closing an aggressor DRAM row, while RowPress induces bitflips by keeping an aggressor DRAM row open for a long period of time. In this {study}, we characterize a DRAM access pattern that combines RowHammer and RowPress {in} 84 real DDR4 DRAM chips from all three major DRAM manufacturers. Our key results {show that 1)} this combined RowHammer and RowPress pattern {takes significantly smaller amount of time (up to 46.1\% faster) to induce the first bitflip compared to the {state-of-the-art} RowPress pattern}, and 2) at the minimum aggressor row activation count to induce at least one bitflip, the bits that flip are different {across} RowHammer, RowPress, and the combined patterns. Based on {our} results, we provide a key hypothesis that the read disturbance effect caused by RowPress from one of the two aggressor rows in a double-sided pattern is much more significant than the other.   
\end{abstract}

\glsresetall
\glsresetall
\section{Introduction}
\label{sec:introduction}
Memory isolation is a fundamental property for system {robustness (i.e., reliability, security, safety).} Accesses to one memory location should \emph{not} induce unintended side-effects on other {(\emph{unaccessed})} memory locations. Unfortunately, the prevalent main memory technology, dynamic random access memory (DRAM)~\cite{dennard1968dram}, is vulnerable to \emph{read disturbance} (i.e., accessing a DRAM cell disturbs the integrity of data in physically adjacent but unaccessed DRAM cells) that can violate memory isolation. 

RowHammer\om{~\cite{kim2014flipping, mutlu2017rowhammer, mutlu2019rowhammer, kim2020revisiting, orosa2021deeper, yaglikci2022understanding, mutlu2023fundamentally, mutlu2023retrospective, olgun2023hbm2disrupt, olgun2024hbm2}} and RowPress~\cite{luo2023rowpress} are two read disturbance phenomenon identified and demonstrated in commodity DRAM chips. RowHammer causes \emph{bitflips} in a DRAM row (victim row) by repeatedly opening and closing \atb{a} physically adjacent DRAM row (aggressor row). RowPress causes bitflips in the victim row by keeping the aggressor row open for a long period of time (i.e., having a longer aggressor row on time, \DRAMTIMING{AggON}). Prior works~\cite{luo2023rowpress, Nam2023XRAY} show that RowHammer and RowPress have \emph{different} underlying read-disturb mechanisms, and cause bitflips with \emph{different} directionalities~\cite{luo2023rowpress, Nam2023XRAY}.

Read disturbance is a critical vulnerability because attackers can leverage bitflips {to perform privilege escalation and leak data}~\exploitingRowHammerAllCitationsExceptFlipping{}. For {robust (i.e., secure, safe, and reliable)} operation of computing systems, it is critical to develop a comprehensive understanding of DRAM read disturbance.

\begin{figure}[h]
    \centering
    \includegraphics[width=0.95\linewidth]{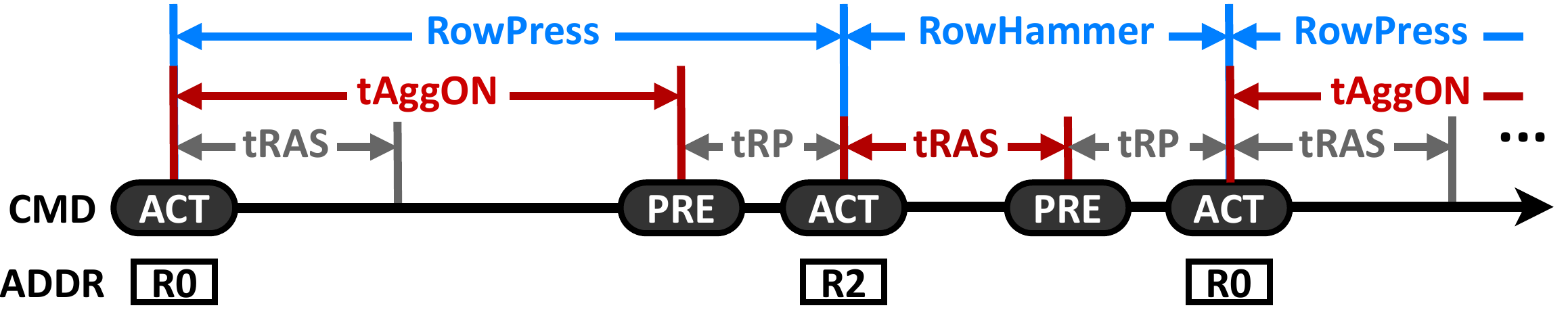}
    \caption{The combined RowHammer and RowPress pattern.}
    \label{fig:pattern_intro}
\end{figure}

In this paper, \textbf{our goal} is to experimentally characterize the bitflips caused by a DRAM access pattern that combines {both} RowHammer and RowPress. As Fig.~\ref{fig:pattern_intro} shows, this involves repeated activations of two aggressor rows (i.e., R0 and R2), where one aggressor row {(R2)} is open for only the minimal amount of time specified by the DRAM standard (i.e., RowHammer, \DRAMTIMING{AggON} $=$ \DRAMTIMING{RAS}), while the other aggressor row {(R0)} is open for a longer period of time (i.e., RowPress, \DRAMTIMING{AggON} $>$ \DRAMTIMING{RAS}). 

We characterize this combined pattern on 84 commodity DDR4 DRAM chips {(from 14 DRAM modules)} from all three major DRAM manufacturers and spans different die densities and die revisions. Our key characterization results demonstrate that 1) the combined pattern induces bitflips faster {(up to 46.1\%)} than conventional RowPress patterns and with much {fewer} {(up to 46.9\%)} aggressor row activations than conventional RowHammer patterns, and 2) induces different bitflips as \DRAMTIMING{AggON} increases compared to conventional RowPress and RowHammer patterns. Based on {our experimental results}, {we hypothesize} that the read disturbance effect caused by RowPress from one of the two aggressor rows in a double-sided pattern is much more significant than the other.

We make the following contributions in this paper:
\begin{itemize}
    \item To our knowledge, this is the first work to experimentally characterize the bitflips from a combined RowHammer and RowPress access pattern in {real} DRAM chips.
    \item We demonstrate the key differences of the bitflips caused by the combined RowHammer and RowPress pattern and conventional RowPress {and RowHammer} patterns.
    \item {We} provide insights {into} and hypotheses {about} the low-level failure mechanisms of RowHammer and RowPress.
\end{itemize}

\glsresetall
\section{Background}
\label{sec:background}

\subsection{DRAM Organization}
\label{sec:dram_org}
\begin{figure}[h]
    \centering
    \includegraphics[width=\linewidth]{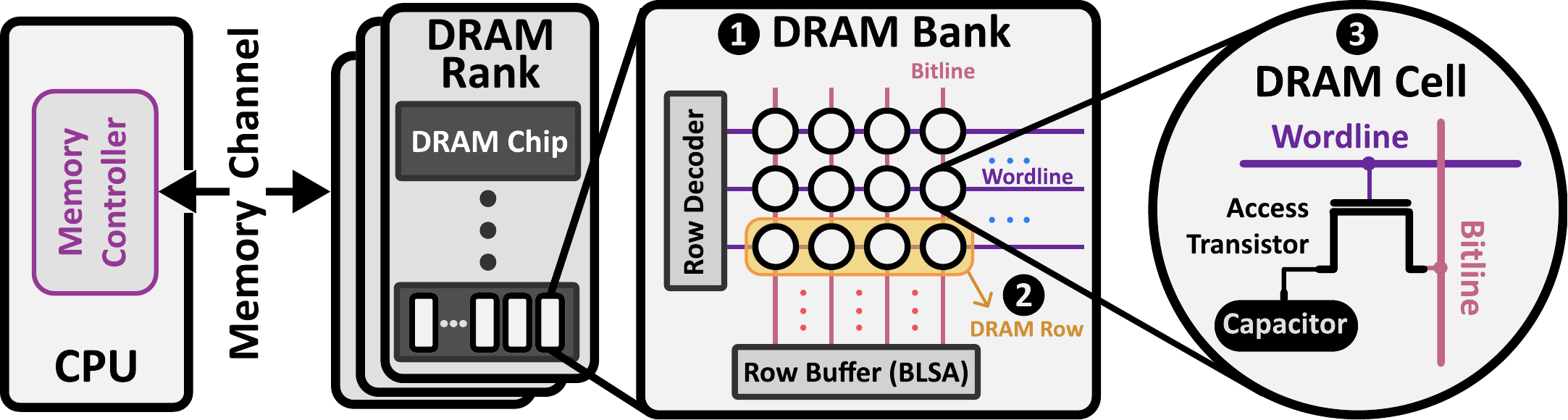}
    \caption{Hierarchical organization of modern DRAM. {Reproduced from~\cite{yaglikci2022understanding, luo2023rowpress}.}}
    \label{fig:dram_org}

\end{figure}

\figref{fig:dram_org} illustrates the hierarchical organization of DRAM-based main memory. A \emph{memory controller} communicates with one or more \emph{memory ranks} over a \emph{memory channel}. A rank consists of multiple DRAM \emph{chips} that operates {in} lock-step. Inside a DRAM chip, there are multiple DRAM \emph{banks} \dingOne{} that can be accessed independently. In a bank, multiple DRAM \emph{cells} \dingThree{} are organized into a 2D array. A DRAM cell stores one bit of information in the form of {electrical} charge in the \emph{capacitor}, connected to a \emph{bitline} through an \emph{access transistor} controlled by a \emph{wordline}.

 DRAM cells are accessed at \emph{row} \dingTwo{} granularity. When {activated}, the wordline of a row is driven high, enabling all the access transistors of the DRAM cells in the row that connect the capacitors to their respective bitlines. The bitlines are connected to the row buffer, which is used to read from or write to the DRAM cells.
\subsection{Key DRAM Operation and Timing}
\label{sec:dram_op}
\noindent\textbf{DRAM Access.} To access DRAM, the memory controller first sends an Activate (\DRAMCMD{ACT}) command to a DRAM bank, {which \emph{opens}} a DRAM row {(i.e., places {the row's} contents into the row buffer)}. {Second, the memory controller} sends Read/Write commands (\DRAMCMD{RD}/\DRAMCMD{WR}) to the DRAM cells in the opened row. {Third}, to access another row in the same bank, the memory controller sends a Precharge  (\DRAMCMD{PRE}) command {that} \emph{closes} the currently open row. The DRAM row must {remain} open (closed) for at least \DRAMTIMING{RAS} (\DRAMTIMING{RP}) amount of time.

\noindent\textbf{DRAM Refresh.} DRAM needs to be \emph{periodically refreshed} to ensure data integrity because the capacitors in DRAM cells lose charge over time{~\cite{liu2012raidr, liu2013experimental}}. Therefore, the memory controller periodically ({every} \DRAMTIMING{REFI}) sends refresh (\DRAMCMD{REF}) commands to DRAM to {restore lost charge}. The JEDEC DDR4 standard~\cite{jedec2017ddr4} specifies that $\text{\DRAMTIMING{REFI}} = 7.8\mu s$ and each DRAM row must be refreshed once every $\text{\DRAMTIMING{REFW}} = 64ms$ under normal operating conditions.
\subsection{DRAM Read Disturbance}
\label{sec:dram_readdisturbance}
DRAM read disturbance is {the} phenomenon that accessing a DRAM row (i.e., the aggressor row) disturbs the charge stored in the DRAM cells in physically adjacent {(\emph{unaccessed})} DRAM rows (i.e., victim rows), causing \emph{bitflips}.

\noindent\textbf{RowHammer.} RowHammer causes bitflips in victim rows through many (e.g., tens of thousands of) repeated {openings and closings (activation \& precharge)} of the aggressor row\om{~\cite{kim2014flipping, mutlu2017rowhammer, mutlu2019rowhammer, kim2020revisiting, orosa2021deeper, yaglikci2022understanding, mutlu2023fundamentally, mutlu2023retrospective, olgun2023hbm2disrupt, olgun2024hbm2}}. In a RowHammer access pattern, each aggressor row is opened (closed) for the minimum amount of time {specified by the DRAM standard} (i.e., aggressor row on time $\text{\DRAMTIMING{AggON}} = \text{\DRAMTIMING{RAS}}$, and aggressor row off time $\text{\DRAMTIMING{AggOFF}} = \text{\DRAMTIMING{RP}}$).

\noindent\textbf{RowPress.} RowPress causes bitflips in victim rows by keeping the aggressor row open for a long period of time  (i.e., $\DRAMTIMING{AggON} > \DRAMTIMING{RAS}$){~\cite{luo2023rowpress}}. Compared to RowHammer bitflips, as \DRAMTIMING{AggON} increases, RowPress bitflips require (much) \emph{fewer} aggressor row activations to induce and have an \emph{opposite} direction {compared to RowHammer}~\cite{luo2023rowpress}.

\section{Experimental Methodology}
\label{sec:methodology}

\subsection{DRAM Characterization Infrastructure}
\label{sec:infra}

We develop {an} FPGA-based commodity DRAM chip characterization infrastructure {building on} DRAM Bender~\cite{olgun2022drambender, drambendergithub} and SoftMC~\cite{hassan2017softmc, softmcgithub}. The infrastructure enables 1) fine-grained control over the DRAM commands and timings, and 2) stable temperature control\footnote{The maximum variation in temperature readings we observe over 24 hours is $\pm$0.2$^{\circ}$C from the target temperature.} of the {tested} DRAM chips tested with heater pads controlled by a PID-based temperature controller~\cite{maxwellFT200}.

We avoid potential interference to  directly observe and analyze the bitflips from the circuit-level following a similar methodology {used} in prior works\om{~\cite{kim2020revisiting, orosa2021deeper, yaglikci2022understanding, hassan2021utrr, luo2023rowpress, olgun2023hbm2disrupt, olgun2024hbm2}}. First, we do \emph{not} send periodic \DRAMCMD{REF} commands to  the DRAM under test to 1) keep the timings of our experiments precise, and 2) not trigger any on-die RowHammer mitigation mechanisms (e.g., target-row-refresh, TRR~\cite{frigo2020trrespass, hassan2021utrr}). Second, we make sure the runtime of each iteration of our characterization experiment does not exceed $60ms$ (strictly smaller than $\text{\DRAMTIMING{REFW}} = 64ms$) to avoid any retention failure bitflips. Third, we do not implement rank-level ECC in our infrastructure and make sure that the DRAM chips we test do \emph{not} have on-die ECC. 

\subsection{Commodity DDR4 DRAM Chips Tested}
\label{sec:dram_tested}
Table~\ref{tab:dram_chip_list} describes the 84 (14) DRAM chips (modules) we test from all three major DRAM manufacturers (Mfr. S, H, and M). For each manufacturer, we test a variety of DRAM die densities and revisions. To account for row address remapping inside DRAM, we reverse-engineer the physical layout of the DRAM rows, following prior works' methodology\om{~\cite{kim2020revisiting, orosa2021deeper, yaglikci2022understanding, hassan2021utrr, luo2023rowpress, olgun2023hbm2disrupt, olgun2024hbm2}}.
\begin{table}[h!]
  \renewcommand{\arraystretch}{0.9}
  \centering
  \scriptsize
  \captionsetup{justification=centering, singlelinecheck=false, labelsep=colon}
\vspace{.1em}
  \caption{DDR4 DRAM Chips Tested.}
  \vspace{-.5em}
  \setlength\tabcolsep{6pt}
    \begin{tabular}{ccccccc}
        \toprule
            {{\bf Mfr.}} & \textbf{\#DIMMs} & {{\bf  \#Chips}}  & {{\bf Density}} & {{\bf Die Rev.}}& {{\bf Org.}}& {{\bf Date}}\\
        \midrule
\multirow{3}{4em}{Mfr. S \\ (Samsung)} & 1 & 8  & 8Gb  & C   & x8  & N/A   \\                         
                                       & 3 & 24 & 8Gb  & D   & x8  & 2110  \\                           
                                       & 1 & 8  & 16Gb  & A  & x8  & 2212   \\                         
        \midrule
\multirow{2}{4em}{Mfr. H \\ (Hynix)}  & 2 & 8  & 8Gb  & D   & x8  & Mar. 21   \\               
                                         & 2 & 8  & 16Gb & C   & x8  & 2136  \\                                                  
     \midrule                  
\multirow{5}{4em}{Mfr. M \\ (Micron)}     
                                        & 1 & 4  & 4Gb  & F   & x16 & N/A  \\                            
                                        & 2 & 16 & 8Gb  & B   & x8  & N/A  \\     
                                        & 1 & 4  & 16Gb & B   & x16 & 2126  \\ 
                                        & 1 & 4  & 16Gb & E   & x16 & 2046  \\                            
        \bottomrule
    \end{tabular}
    \label{tab:dram_chip_list}
\end{table}
\subsection{Combined RowHammer and RowPress Pattern}
\label{sec:pattern}

\begin{figure}[h]
\centering
\includegraphics[width=\linewidth]{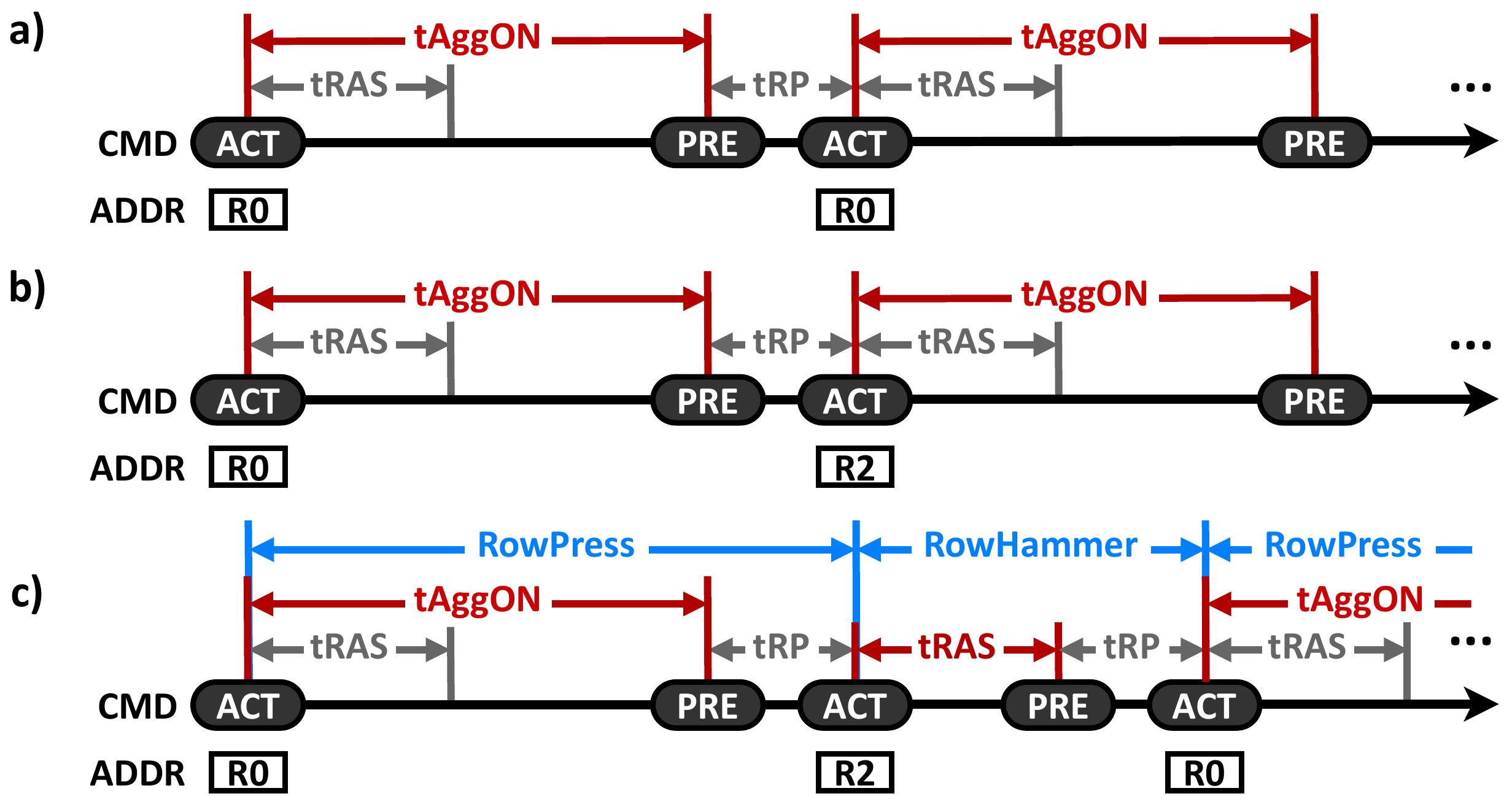}
\caption{Comparison of a) the conventional single-sided RowPress (RowHammer, when \DRAMTIMING{AggON} = \DRAMTIMING{RAS}) pattern, b) the conventional double-sided RowPress (RowHammer, when \DRAMTIMING{AggON} = \DRAMTIMING{RAS}) pattern, and c) the combined RowHammer and RowPress pattern.}
\label{fig:pattern}
\end{figure}

\figref{fig:pattern} shows command sequences and timings of the DRAM access patterns we characterize in this paper. \figref{fig:pattern}.a shows the conventional single-sided RowPress pattern involving only one aggressor row (R0) that is open for \DRAMTIMING{AggON} amount of time per activation. If \DRAMTIMING{AggON} $=$ \DRAMTIMING{RAS}, then this pattern is identical to the conventional single-sided RowHammer pattern. \figref{fig:pattern}.b shows the conventional double-sided RowPress pattern involving alternating activations to two aggressor rows (R0 and R2). Both R0 and R2 are open for \DRAMTIMING{AggON} per activation. When \DRAMTIMING{AggON} $=$ \DRAMTIMING{RAS}, this pattern is identical to the conventional double-sided RowHammer pattern. \figref{fig:pattern}.c shows the combined RowHammer and RowPress access pattern {(that is not explored in prior work)}. This pattern involves alternating activations to two aggressor rows (R0 and R2), but R0 is open for \DRAMTIMING{AggON} ($> \text{\DRAMTIMING{RAS}}$) amount of time and R2 is \emph{always} open for \DRAMTIMING{RAS}, the minimal amount of row open time allowed by the JEDEC standard.

\label{sec:exp_methodology}
\subsection{Real DRAM Chip Characterization Methodology}
For each DRAM module, we evaluate the test patterns on 3K DRAM rows in an arbitrarily chosen DRAM bank (1K rows at the beginning, middle, and end of the bank, respectively). We use a checkerboard data pattern that initializes the aggressor row(s) with \texttt{0xAA} and the victim row(s) with \texttt{0x55}. For each pattern, we sweep \DRAMTIMING{AggON} from the minimum value of 36ns (i.e., \DRAMTIMING{AggON} $=$ \DRAMTIMING{RAS}) up to $300\mu s$.  We repeat each experiment to measure \gls{acmin} three times. We conduct all our characterization at $50^{\circ}C$.
\section{Major Characterization Results}
\label{sec:majpr_results}

\begin{figure}[!h]
\centering
\includegraphics[width=0.95\linewidth]{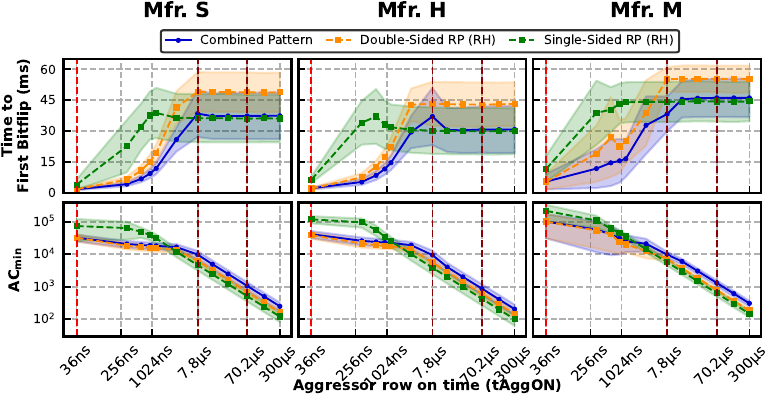}
\caption{Time to first bitflip (first row of plots) and \gls{acmin} (second row of plots) of the combined RowHammer and RowPress pattern (blue solid line) and the conventional single- and double-sided RowPress (RowHammer) patterns (green and orange dashed lines).}
\label{fig:agg_acmin_ttfb}
\end{figure}

\figref{fig:agg_acmin_ttfb} shows how time to first bitflip (y-axis, first row of plots) and \gls{acmin} (y-axis, second row of plots) of the combined RowHammer and RowPress pattern (solid blue lines) and the conventional double-sided RowPress (RowHammer) pattern (dashed orange lines) changes as \DRAMTIMING{AggON} (x-axis) increases for DRAM {modules} from Mfr. S, H, and M, respectively, at $50^{\circ}C$. Each data point shows the average time to first bitflip or \gls{acmin} at a given \DRAMTIMING{AggON} value across all tested DRAM dies for each manufacturer. The error band represents the standard deviation. We highlight \DRAMTIMING{AggON} $=$ 36ns ($=$ \DRAMTIMING{RAS}) as dashed {dark red} lines on the x-axis because both the combined pattern and the conventional double-sided RowPress pattern are identical to the conventional double-sided RowHammer pattern when \DRAMTIMING{AggON} $=$ 36ns ($=$ \DRAMTIMING{RAS}). We highlight \DRAMTIMING{AggON} $=$ \SI{7.8}{\micro\second} ($=$ \DRAMTIMING{REFI}) and \SI{70.2}{\micro\second} ($=9\times$\DRAMTIMING{REFI}) as dashed dark red lines on the x-axis because these are the potential upper bounds of \DRAMTIMING{AggON} as specified by the JEDEC standard~\cite{jedec2017ddr4}. We make three major observations from \figref{fig:agg_acmin_ttfb}.

\observation{As \DRAMTIMING{AggON} initially starts to increase, the combined RowHammer and RowPress pattern takes much less time to induce the first bitflip compared to both the conventional single- and double-sided RowPress patterns.}

For example, when $\text{\DRAMTIMING{AggON}}=636ns$, it takes the combined pattern only \SI{6.8}{\milli\second}, \SI{8.5}{\milli\second}, \SI{14.6}{\milli\second} on average to induce the first bitflip in the victim row, for Mfr. S, H, M, respectively. This is $37.6\%$, $33.6\%$, $46.1\%$ faster compared to the conventional double-sided RowPress pattern (which takes \SI{10.9}{\milli\second}, \SI{12.8}{\milli\second}, \SI{27.1}{\milli\second} for Mfr. S, H, M, respectively, to induce the first bitflip). Compared to the single-sided RowPress pattern (which takes \SI{32.2}{\milli\second}, \SI{37.1}{\milli\second}, \SI{40.4}{\milli\second} for Mfr. S, H, M, respectively, to induce the first bitflip), the combined pattern is $78.9\%$, $77.1\%$, $63.9\%$ faster.

\takeawaybox{Read disturbance bitflips can be induced in a {smaller} amount of time by combining RowPress and RowHammer compared to {using solely} RowPress or RowHammer.}

We hypothesize that the reason for Observations 1 is that in a double-sided RowPress pattern, the read disturbance effect caused by RowPress from one of the two aggressor rows is much more significant compared to the other such that reducing the \DRAMTIMING{AggON} of this other aggressor row does \emph{not} significantly change \gls{acmin}.

\hypobox{As \DRAMTIMING{AggON} initially starts to increase, the read disturbance effect caused by RowPress from one of the two aggressor rows in the double-sided pattern is much more significant than the other.}

\observation{As \DRAMTIMING{AggON} initially starts to increase, the combined pattern needs slightly more aggressor row activations to induce at least {one} bitflip than the conventional double-sided RowPress pattern.}

When $\text{\DRAMTIMING{AggON}}=636ns$, compared to $\text{\DRAMTIMING{AggON}}=36ns$ (i.e., RowHammer), the \gls{acmin} of the combined pattern reduces by 40.5\%, 42.0\%, 46.9\% on average for Mfr. S, H, M, respectively. {This is 7.5\%, 8.0\%, and 7.4\% less \gls{acmin} reduction for Mfr. S, H, M, respectively, compared to the conventional double-sided RowPress pattern (48.0\%, 50.0\%, 54.3\%)}.

\observation{As \DRAMTIMING{AggON} continues to increase, the combined pattern takes a similar amount of time to induce the first bitflip as the conventional single-sided RowPress pattern.}

When $\text{\DRAMTIMING{AggON}} = 70.2\mu s$, the combined pattern takes on average 37.4ms, 30.8ms, 46.1ms to induce the first bitflip for Mfr. S, H, and M, respectively. This is 3.9\%, 3.0\%, 4.1\% slower than the conventional single-sided RowPress pattern, which takes 36.0ms, 29.9ms, 44.3ms to induce the first bitflip. 

We hypothesize that the reason for Observation 3 is that as \DRAMTIMING{AggON} becomes large, the read disturbance effect from RowPress is dominant {compared to RowHammer} due to the significantly reduced number of aggressor row activations, causing the combined RowHammer and RowPress pattern to behave very similarly to {the} conventional single-sided RowPress pattern.

\hypobox{For large \DRAMTIMING{AggON} values, the read disturbance effect from RowPress is dominant {compared to RowHammer} in the combined RowHammer and RowPress pattern.}
\pagebreak

\begin{figure}[h]
\centering
\includegraphics[width=\linewidth]{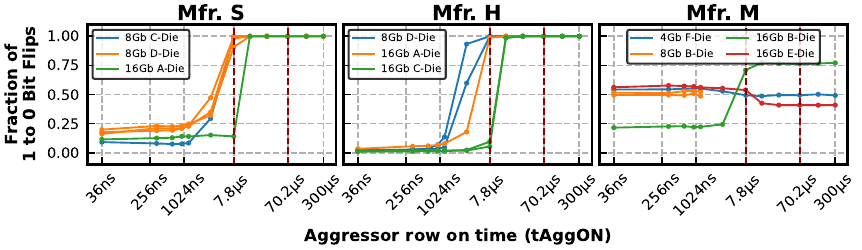}
\caption{The fraction of 1 to 0 bitflips {due to} the combined RowHammer and RowPress pattern.}
\label{fig:direction}
\end{figure}

\figref{fig:direction} shows the fraction of 1-to-0 bitflips of all the bitflips we observe from the combined RowHammer and RowPress pattern. We make the following observation from \figref{fig:direction}.

\observation{As \DRAMTIMING{AggON} increases, the directionality of bitflips caused by the combined RowHammer and RowPress pattern changes.}

We observe that for all DRAM dies tested from Mfr. S and H, as \DRAMTIMING{AggON} initially starts to increase, the majority of the bitflips from the combined pattern are 0-to-1 bitflips. As \DRAMTIMING{AggON} continues to increase, the fraction of 1-to-0 bitflips significantly increases. For sufficiently large \DRAMTIMING{AggON} values, almost 100\% of the bitflips are 1-to-0. Such {a} change in the directionality of bitflips as \DRAMTIMING{AggON} increases is the same {observation} as \om{in} {the original RowPress paper}~\cite{luo2023rowpress}.\footnote{For Mfr. M, we observe an opposite trend where the fraction of 1-to-0 bitflips \emph{decreases} as \DRAMTIMING{AggON} increases for all but the 16Gb B-Dies. We hypothesize that this is a result of a different true- and anti-cell layout in Mfr. M's DRAM design compared to the other two manufacturers. Such an observation on DRAM dies from Mfr. M is similar to that {in the original RowPress paper}\cite{luo2023rowpress}.} This observation also supports our Hypothesis 2 that the RowPress effect is dominant in the combined RowHammer and RowPress pattern.

\begin{figure}[h]
\centering
\includegraphics[width=\linewidth]{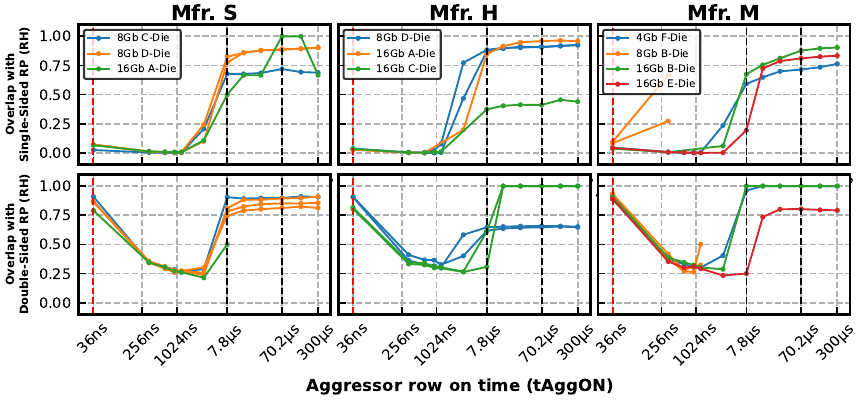}
\caption{The overlap ratio between the bitflips from the combined {RowHammer and RowPress} pattern and the conventional single-sided (top row of plots) and double-sided (bottom row of plots) RowPress (RowHammer) pattern.}
\label{fig:overlap_ds}
\end{figure}

\figref{fig:overlap_ds} shows the overlap (y-axis) between the bitflips from the combined RowHammer and RowPress pattern and the conventional single- (first row of plots) and double-sided (second row of plots) RowPress (RowHammer) pattern as \DRAMTIMING{AggON} (x-axis) increases. We define such overlap as the number of unique bitflips that are observed in both the combined pattern and the conventional RowPress (RowHammer) patterns divided by the total number of unique bitflips {observed in} the conventional pattern. We make two observations from the figure.

\observation{The overlap between the bitflips from the combined pattern and conventional single-sided RowPress pattern increases as \DRAMTIMING{AggON} increases.}

\observation{The overlap between the bitflips from the combined pattern and conventional double-sided RowPress pattern first decreases as \DRAMTIMING{AggON} initially starts to increase, and then increases as \DRAMTIMING{AggON} continues to increase.}

When \DRAMTIMING{AggON} initially starts to increase, the overlap between the bitflips from the combined RowHammer and RowPress pattern and the conventional single-sided RowPress pattern remains very small, but the overlap between the bitflips from the combined pattern and the conventional double-sided RowPress pattern significantly decreases. As \DRAMTIMING{AggON} continues to increase beyond a certain level (e.g., $>$ \SI{7.8}{\micro\second}), both the overlap between the combined RowHammer and RowPress pattern and the conventional single- (double-) sided RowPress patterns significantly increases to more than 75\%. 

\takeawaybox{The combined RowHammer and RowPress pattern induces different bitflips compared to the conventional single- and double-sided RowPress patterns.}
\section{Related Works}
\label{sec:related_works}
To our knowledge, this is the first work to experimentally demonstrate and characterize read disturbance caused by a combined RowHammer and RowPress access pattern. Existing works on experimental characterization of DRAM read disturbance test either {only} RowHammer patterns~\cite{kim2014flipping, kim2020revisiting, orosa2021deeper, yaglikci2022understanding} or {separate RowHammer and RowPress patterns}~\cite{luo2023rowpress, Nam2023XRAY, olgun2023hbm2disrupt, yaglikci2024svard, olgun2024hbm2} patterns. {Prior works on device-level mechanisms of RowHammer~\cite{yang2016suppression, park2016experiments, ryu2017overcoming, yang2019trap, walker2021ondramrowhammer, zhou2023Double, jie2024understanding} {and RowPress~\cite{zhou2024understanding}} do not investigate \om{combining} RowHammer and RowPress.}

\section{Conclusion}
In this paper, we experimentally demonstrate and characterize, for the first time, the bitflips caused by a DRAM access pattern that combines RowHammer and RowPress. Our characterization results show that the combined access pattern 1) induces bitflips faster compared to conventional single- and double-sided RowPress patterns, and 2) induces different bitflips compared to single- and double-sided RowPress (RowHammer) patterns. 

We plan to investigate deeper into the combined RowHammer and RowPress pattern by 1) performing more comprehensive and rigorous characterization and analysis of {the bitflips by testing more DRAM chips with more data patterns and temperatures,} 2) look into the device-level mechanisms of RowHammer and RowPress to verify our hypotheses, and 3) understand {the} architectural implications by analyzing and evaluating how existing mitigation mechanisms need to be changed.

We hope the results and insights from this paper lead to more comprehensive and fundamental understanding of DRAM read disturbance and further research in building {more robust} DRAM-based memory systems.
\section*{\ieycr{0}{{Acknowledgments}}}
\ieycr{0}{We thank the anonymous reviewers of DSN Disrupt 2024 for their encouraging feedback. 
We thank the SAFARI Research Group members for providing a stimulating intellectual environment. We acknowledge the generous gifts from our industrial partners, including Google, Huawei, Intel, and Microsoft. This work is supported in part by the {Microsoft-Swiss Joint Research Center and a Google Security \& Privacy Research Award}.}

\balance 
{
  \bstctlcite{IEEEexample:BSTcontrol}
  \let\OLDthebibliography\thebibliography
  \renewcommand\thebibliography[1]{
    \OLDthebibliography{#1}
    \setlength{\parskip}{0pt}
    \setlength{\itemsep}{0pt}
  }
  \bibliographystyle{IEEEtran}
  \bibliography{refs}
}

\doublespacing
\clearpage
\onecolumn
\begin{landscape}
\section{\textbf{Appendix}}
\label{sec:appendix}

\begin{table}[h]
\centering
\caption{{Summary of all tested DDR4 modules and their vulnerabilities to conventional double-sided RowHammer/RowPress patterns and the combined RH/RP pattern {in terms of \gls{acmin} (i.e., the minimum number of total aggressor row activations needed to induce at least one bitflip) and the time to the first bitflip.}}}
\label{tab:extended-module-info}
\resizebox{\columnwidth}{!}{%
\begin{tabular}{@{}cccccc|c|ccccc|ccccc|@{}}
\toprule
 &
   &
   &
   &
   &
   &
   &
  \multicolumn{5}{c|}{\textbf{\begin{tabular}[c]{@{}c@{}}ACmin @ Different tAggON, Double-Sided\\ Avg. (Min.)\end{tabular}}} &
  \multicolumn{5}{c|}{\textbf{\begin{tabular}[c]{@{}c@{}}Time to First Bitflip (ms) @ Different tAggON, Double-Sided\\ Avg. (Min.)\end{tabular}}} \\ \cmidrule(l){8-17} 
 &
   &
   &
   &
   &
   &
   &
  \multicolumn{1}{c|}{\textbf{RowHammer (RH)}} &
  \multicolumn{2}{c|}{\textbf{RowPress (RP)}} &
  \multicolumn{2}{c|}{\textbf{Combined RH/RP}} &
  \multicolumn{1}{c|}{\textbf{RowHammer (RH)}} &
  \multicolumn{2}{c|}{\textbf{RowPress (RP)}} &
  \multicolumn{2}{c|}{\textbf{Combined RH/RP}} \\ \cmidrule(l){8-17} 
\multirow{-3}{*}{\textbf{Mfr.}} &
  \multirow{-3}{*}{\textbf{DIMM Part}} &
  \multirow{-3}{*}{\textbf{DRAM Part}} &
  \multirow{-3}{*}{\textbf{\begin{tabular}[c]{@{}c@{}}Die\\ Rev.\end{tabular}}} &
  \multirow{-3}{*}{\textbf{\begin{tabular}[c]{@{}c@{}}Die\\ Density\end{tabular}}} &
  \multirow{-3}{*}{\textbf{Date}} &
  \multirow{-3}{*}{\textbf{ID}} &
  \multicolumn{1}{c|}{\textbf{\begin{tabular}[c]{@{}c@{}}36ns\\ (tRAS)\end{tabular}}} &
  \multicolumn{1}{c|}{\textbf{\begin{tabular}[c]{@{}c@{}}7.8us\\ (tREFI)\end{tabular}}} &
  \multicolumn{1}{c|}{\textbf{\begin{tabular}[c]{@{}c@{}}70.2us\\ (9xtREFI)\end{tabular}}} &
  \multicolumn{1}{c|}{\textbf{\begin{tabular}[c]{@{}c@{}}7.8us\\ (tREFI)\end{tabular}}} &
  \textbf{\begin{tabular}[c]{@{}c@{}}70.2us\\ (9xtREFI)\end{tabular}} &
  \multicolumn{1}{c|}{\textbf{\begin{tabular}[c]{@{}c@{}}36ns\\ (tRAS)\end{tabular}}} &
  \multicolumn{1}{c|}{\textbf{\begin{tabular}[c]{@{}c@{}}7.8us\\ (tREFI)\end{tabular}}} &
  \multicolumn{1}{c|}{\textbf{\begin{tabular}[c]{@{}c@{}}70.2us\\ (9xtREFI)\end{tabular}}} &
  \multicolumn{1}{c|}{\textbf{\begin{tabular}[c]{@{}c@{}}7.8us\\ (tREFI)\end{tabular}}} &
  \textbf{\begin{tabular}[c]{@{}c@{}}70.2us\\ (9xtREFI)\end{tabular}} \\ \midrule
 &
  M393A2K40CB2-CTD &
  K4A8G045WC-BCTD &
  C &
  8 Gb &
  2135 &
  \cellcolor[HTML]{DAE8FC}S0 &
  \multicolumn{1}{c|}{\cellcolor[HTML]{DAE8FC}45.0K (22.6K)} &
  \multicolumn{1}{c|}{\cellcolor[HTML]{DAE8FC}6.9K (2.9K)} &
  \multicolumn{1}{c|}{\cellcolor[HTML]{DAE8FC}762 (316)} &
  \multicolumn{1}{c|}{\cellcolor[HTML]{DAE8FC}11.4K (3.2K)} &
  \cellcolor[HTML]{DAE8FC}1.3K (354) &
  \multicolumn{1}{c|}{\cellcolor[HTML]{DAE8FC}2.4 (1.2)} &
  \multicolumn{1}{c|}{\cellcolor[HTML]{DAE8FC}53.8 (22.7)} &
  \multicolumn{1}{c|}{\cellcolor[HTML]{DAE8FC}53.5 (22.2)} &
  \multicolumn{1}{c|}{\cellcolor[HTML]{DAE8FC}44.8 (12.6)} &
  \cellcolor[HTML]{DAE8FC}45.6 (12.4) \\ \cmidrule(l){2-17} 
 &
   &
   &
   &
   &
   &
  S1 &
  \multicolumn{1}{c|}{28.6K (16.2K)} &
  \multicolumn{1}{c|}{6.7K (2.5K)} &
  \multicolumn{1}{c|}{739 (280)} &
  \multicolumn{1}{c|}{10.3K (2.5K)} &
  1.2K (292) &
  \multicolumn{1}{c|}{1.6 (0.9)} &
  \multicolumn{1}{c|}{52.4 (19.2)} &
  \multicolumn{1}{c|}{51.8 (19.7)} &
  \multicolumn{1}{c|}{40.5 (9.7)} &
  41.2 (10.3) \\
 &
   &
   &
   &
   &
   &
  \cellcolor[HTML]{DAE8FC}S2 &
  \multicolumn{1}{c|}{\cellcolor[HTML]{DAE8FC}28.8K (16.0K)} &
  \multicolumn{1}{c|}{\cellcolor[HTML]{DAE8FC}5.8K (1.6K)} &
  \multicolumn{1}{c|}{\cellcolor[HTML]{DAE8FC}648 (180)} &
  \multicolumn{1}{c|}{\cellcolor[HTML]{DAE8FC}7.2K (1.6K)} &
  \cellcolor[HTML]{DAE8FC}798 (184) &
  \multicolumn{1}{c|}{\cellcolor[HTML]{DAE8FC}1.6 (0.9)} &
  \multicolumn{1}{c|}{\cellcolor[HTML]{DAE8FC}45.5 (12.3)} &
  \multicolumn{1}{c|}{\cellcolor[HTML]{DAE8FC}45.5 (12.6)} &
  \multicolumn{1}{c|}{\cellcolor[HTML]{DAE8FC}28.2 (6.4)} &
  \cellcolor[HTML]{DAE8FC}28.0 (6.5) \\
 &
  \multirow{-3}{*}{M378A1K43DB2-CTD} &
  \multirow{-3}{*}{K4A8G085WD-BCTD} &
  \multirow{-3}{*}{D} &
  \multirow{-3}{*}{8 Gb} &
  \multirow{-3}{*}{2110} &
  S3 &
  \multicolumn{1}{c|}{29.2K (15.8K)} &
  \multicolumn{1}{c|}{6.5K (1.6K)} &
  \multicolumn{1}{c|}{717 (186)} &
  \multicolumn{1}{c|}{9.0K (1.6K)} &
  1.0K (174) &
  \multicolumn{1}{c|}{1.6 (0.9)} &
  \multicolumn{1}{c|}{50.5 (12.8)} &
  \multicolumn{1}{c|}{50.3 (13.0)} &
  \multicolumn{1}{c|}{35.2 (6.4)} &
  35.3 (6.1) \\ \cmidrule(l){2-17} 
\multirow{-5}{*}{\textbf{\begin{tabular}[c]{@{}c@{}}Samsung\\ (Mfr. S)\end{tabular}}} &
  M471A4G43AB1-CWE &
  K4AAG085WA-BCWE &
  A &
  16 Gb &
  2320 &
  \cellcolor[HTML]{DAE8FC}S4 &
  \multicolumn{1}{c|}{\cellcolor[HTML]{DAE8FC}31.3K (17.0K)} &
  \multicolumn{1}{c|}{\cellcolor[HTML]{DAE8FC}7.6K (7.5K)} &
  \multicolumn{1}{c|}{\cellcolor[HTML]{DAE8FC}No Bitflip} &
  \multicolumn{1}{c|}{\cellcolor[HTML]{DAE8FC}14.0K (9.4K)} &
  \cellcolor[HTML]{DAE8FC}1.5K (1.5K) &
  \multicolumn{1}{c|}{\cellcolor[HTML]{DAE8FC}1.7 (0.9)} &
  \multicolumn{1}{c|}{\cellcolor[HTML]{DAE8FC}59.6 (58.2)} &
  \multicolumn{1}{c|}{\cellcolor[HTML]{DAE8FC}No Bitflip} &
  \multicolumn{1}{c|}{\cellcolor[HTML]{DAE8FC}55.1 (36.9)} &
  \cellcolor[HTML]{DAE8FC}54.4 (51.4) \\ \midrule
 &
   &
   &
   &
   &
   &
  H0 &
  \multicolumn{1}{c|}{43.4K (16.0K)} &
  \multicolumn{1}{c|}{6.5K (3.0K)} &
  \multicolumn{1}{c|}{724 (312)} &
  \multicolumn{1}{c|}{8.2K (3.0K)} &
  935 (324) &
  \multicolumn{1}{c|}{2.3 (0.9)} &
  \multicolumn{1}{c|}{51.0 (23.1)} &
  \multicolumn{1}{c|}{50.8 (21.9)} &
  \multicolumn{1}{c|}{32.3 (11.7)} &
  32.8 (11.4) \\
 &
  \multirow{-2}{*}{\begin{tabular}[c]{@{}c@{}}(Kingston)\\ KSM32RD8/16HDR\end{tabular}} &
  \multirow{-2}{*}{H5AN8G8NDJR-XNC} &
  \multirow{-2}{*}{D} &
  \multirow{-2}{*}{8 Gb} &
  \multirow{-2}{*}{2048} &
  \cellcolor[HTML]{DAE8FC}H1 &
  \multicolumn{1}{c|}{\cellcolor[HTML]{DAE8FC}45.6K (21.4K)} &
  \multicolumn{1}{c|}{\cellcolor[HTML]{DAE8FC}4.7K (1.6K)} &
  \multicolumn{1}{c|}{\cellcolor[HTML]{DAE8FC}509 (170)} &
  \multicolumn{1}{c|}{\cellcolor[HTML]{DAE8FC}6.0K (1.7K)} &
  \cellcolor[HTML]{DAE8FC}646 (184) &
  \multicolumn{1}{c|}{\cellcolor[HTML]{DAE8FC}2.5 (1.2)} &
  \multicolumn{1}{c|}{\cellcolor[HTML]{DAE8FC}36.4 (12.1)} &
  \multicolumn{1}{c|}{\cellcolor[HTML]{DAE8FC}35.8 (11.9)} &
  \multicolumn{1}{c|}{\cellcolor[HTML]{DAE8FC}23.6 (6.7)} &
  \cellcolor[HTML]{DAE8FC}22.7 (6.5) \\ \cmidrule(l){2-17} 
 &
   &
   &
   &
   &
   &
  H2 &
  \multicolumn{1}{c|}{33.1K (15.8K)} &
  \multicolumn{1}{c|}{6.9K (3.5K)} &
  \multicolumn{1}{c|}{699 (376)} &
  \multicolumn{1}{c|}{13.7 (3.5K)} &
  1.5K (386) &
  \multicolumn{1}{c|}{1.8 (0.9)} &
  \multicolumn{1}{c|}{54.1 (27.3)} &
  \multicolumn{1}{c|}{54.8 (20.5)} &
  \multicolumn{1}{c|}{53.6 (13.7)} &
  51.5 (13.6) \\
\multirow{-4}{*}{\textbf{\begin{tabular}[c]{@{}c@{}}SK Hynix\\ (Mfr. H)\end{tabular}}} &
  \multirow{-2}{*}{HMAA4GU6AJR8N-XN} &
  \multirow{-2}{*}{H5ANAG8NAJR-XN} &
  \multirow{-2}{*}{C} &
  \multirow{-2}{*}{16 Gb} &
  \multirow{-2}{*}{2051} &
  \cellcolor[HTML]{DAE8FC}H3 &
  \multicolumn{1}{c|}{\cellcolor[HTML]{DAE8FC}32.9K (15.9K)} &
  \multicolumn{1}{c|}{\cellcolor[HTML]{DAE8FC}7.6K (6.7K)} &
  \multicolumn{1}{c|}{\cellcolor[HTML]{DAE8FC}839 (814)} &
  \multicolumn{1}{c|}{\cellcolor[HTML]{DAE8FC}13.7K (7.0K)} &
  \cellcolor[HTML]{DAE8FC}1.4K (794) &
  \multicolumn{1}{c|}{\cellcolor[HTML]{DAE8FC}1.8 (0.9)} &
  \multicolumn{1}{c|}{\cellcolor[HTML]{DAE8FC}59.5 (52.8)} &
  \multicolumn{1}{c|}{\cellcolor[HTML]{DAE8FC}58.9 (57.1)} &
  \multicolumn{1}{c|}{\cellcolor[HTML]{DAE8FC}53.9 (27.3)} &
  \cellcolor[HTML]{DAE8FC}50.1 (27.9) \\ \midrule
 &
  CT40K512M8SA-075E:F &
  CT4G4DFS8266.C8FF &
  F &
  4 Gb &
  2107 &
  M0 &
  \multicolumn{1}{c|}{71.0K (31.0K)} &
  \multicolumn{1}{c|}{6.9K (3.6K)} &
  \multicolumn{1}{c|}{755 (396)} &
  \multicolumn{1}{c|}{12.7K (3.7K)} &
  1.5K (410) &
  \multicolumn{1}{c|}{3.8 (1.7)} &
  \multicolumn{1}{c|}{53.6 (27.9)} &
  \multicolumn{1}{c|}{53.0 (27.8)} &
  \multicolumn{1}{c|}{49.9 (14.3)} &
  51.0 (14.4) \\ \cmidrule(l){2-17} 
 &
   &
   &
   &
   &
  1911 &
  \cellcolor[HTML]{DAE8FC}M1 &
  \multicolumn{1}{c|}{\cellcolor[HTML]{DAE8FC}192.7K (83.6K)} &
  \multicolumn{1}{c|}{\cellcolor[HTML]{DAE8FC}No Bitflip} &
  \multicolumn{1}{c|}{\cellcolor[HTML]{DAE8FC}No Bitflip} &
  \multicolumn{1}{c|}{\cellcolor[HTML]{DAE8FC}No Bitflip} &
  \cellcolor[HTML]{DAE8FC}No Bitflip &
  \multicolumn{1}{c|}{\cellcolor[HTML]{DAE8FC}10.4 (4.5)} &
  \multicolumn{1}{c|}{\cellcolor[HTML]{DAE8FC}No Bitflip} &
  \multicolumn{1}{c|}{\cellcolor[HTML]{DAE8FC}No Bitflip} &
  \multicolumn{1}{c|}{\cellcolor[HTML]{DAE8FC}No Bitflip} &
  \cellcolor[HTML]{DAE8FC}No Bitflip \\ \cmidrule(lr){6-6}
 &
  \multirow{-2}{*}{MTA18ASF2G72PZ-2G3B1} &
  \multirow{-2}{*}{MT40A2G4WE-083E:B} &
  \multirow{-2}{*}{B} &
  \multirow{-2}{*}{8 Gb} &
  1903 &
  M2 &
  \multicolumn{1}{c|}{170.0K (75.2K)} &
  \multicolumn{1}{c|}{No Bitflip} &
  \multicolumn{1}{c|}{No Bitflip} &
  \multicolumn{1}{c|}{No Bitflip} &
  No Bitflip &
  \multicolumn{1}{c|}{9.2 (4.1)} &
  \multicolumn{1}{c|}{No Bitflip} &
  \multicolumn{1}{c|}{No Bitflip} &
  \multicolumn{1}{c|}{No Bitflip} &
  No Bitflip \\ \cmidrule(l){2-17} 
 &
  MTA4ATF1G64HZ-3G2B2 &
  MT40A1G16RC-062E:B &
  B &
  16 Gb &
  2126 &
  \cellcolor[HTML]{DAE8FC}M3 &
  \multicolumn{1}{c|}{\cellcolor[HTML]{DAE8FC}53.5K (26.0K)} &
  \multicolumn{1}{c|}{\cellcolor[HTML]{DAE8FC}7.6K (7.3K)} &
  \multicolumn{1}{c|}{\cellcolor[HTML]{DAE8FC}833 (802)} &
  \multicolumn{1}{c|}{\cellcolor[HTML]{DAE8FC}13.6K (9.0K)} &
  \cellcolor[HTML]{DAE8FC}1.6K (1.0K) &
  \multicolumn{1}{c|}{\cellcolor[HTML]{DAE8FC}2.9 (1.4)} &
  \multicolumn{1}{c|}{\cellcolor[HTML]{DAE8FC}59.2 (59.3)} &
  \multicolumn{1}{c|}{\cellcolor[HTML]{DAE8FC}58.5 (56.3)} &
  \multicolumn{1}{c|}{\cellcolor[HTML]{DAE8FC}53.4 (35.2)} &
  \cellcolor[HTML]{DAE8FC}54.8 (35.5) \\ \cmidrule(l){2-17} 
\multirow{-5}{*}{\textbf{\begin{tabular}[c]{@{}c@{}}Micron\\ (Mfr. M)\end{tabular}}} &
  MTA4ATF1G64HZ-3G2E1 &
  MT40A1G16KD-062E:E &
  E &
  16 Gb &
  2046 &
  M4 &
  \multicolumn{1}{c|}{20.2K (10.7K)} &
  \multicolumn{1}{c|}{7.1K (2.6K)} &
  \multicolumn{1}{c|}{790 (272)} &
  \multicolumn{1}{c|}{8.9K (2.7K)} &
  1.3K (296) &
  \multicolumn{1}{c|}{1.1 (0.6)} &
  \multicolumn{1}{c|}{55.2 (20.4)} &
  \multicolumn{1}{c|}{55.5 (19.1)} &
  \multicolumn{1}{c|}{34.9 (10.7)} &
  44.3 (10.4) \\ \bottomrule
\end{tabular}%
}
\end{table}
\end{landscape}

\end{document}